# Self-assembled ErSb nanostructures of tunable shape and orientation: Growth and plasmonic properties

Hong Lu[1*], Daniel G. Ouellette[2*], Sascha Preu[3*], Justin D. Watts[1,2], Ben Zaks[2], Peter G. Burke[1], Mark S. Sherwin[2], and Arthur C. Gossard[1]

1. *Materials Department, University of California, Santa Barbara, California, USA*
2. *Department of Physics and the Institute for Terahertz Science and Technology, University of California, Santa Barbara, California, USA*
3. *Chair of Applied Physics, University of Erlangen-Nuremberg, Erlangen, Germany*

**Self-assembled, semimetallic ErSb single crystal nanostructures, grown by molecular beam epitaxy, are embedded within a semiconductor GaSb matrix. Formation, evolution and orientation of a variety of nanostructures, including spherical nanoparticles, elongated nanorods, octagonal shaped nanowires oriented along the surface normal and nanowires oriented in the growth plane, are controlled simply by the Er fraction. The plasmonic properties of the semimetal/semiconductor composites are characterized and quantified by three polarization-resolved spectroscopy techniques, spanning more than three orders of magnitude in frequency from 100 GHz up to 300 THz. The effect of the size, shape and orientation of the nanostructures is characterized by polarization-sensitive response and modeled by a Maxwell-Garnett effective medium theory.**

Low dimensional systems are widely utilized in current nanotechnology for their exceptional properties at small scales. The ability to control the growth, size and geometry of low dimensional systems is a vital step in engineering optical and electronic properties. Many of the synthesis methods for low dimensional structures involve processes that require templates or catalysts. By molecular beam epitaxy (MBE), low dimensional rare earth pnictides[1] can be grown with semiconductors[2,3] and form a coherent semimetal/semiconductor composite[4]. The incorporation of metallic nanostructures into a semiconductor matrix without affecting the matrix crystalline quality provides opportunities to engineer the composite properties for desired applications[5]. Recently, ErAs nanoparticles have been extensively studied with respect to their semimetallic band structure[1,6,7], and their integration with III-As semiconductors has found potential applications in THz photomixers[8,9], Schottky diodes[10], recombination-enhanced n-ErAs-p diodes[11-13], enhanced tunneling for high efficiency multi-junction solar cells[14], and enhanced thermoelectric figure-of-merit for efficient thermoelectric power generation[15,16]. The ErSb:GaSb composite system in many ways is the most ideal semimetal-semiconductor system to date. ErSb is lattice-matched to GaSb within 0.2% (vs. 1.53% for the ErAs:GaAs system) and its band alignment with GaSb allows p-type applications of the composites. However, this material system is still underexplored.

In this manuscript we report growth of a variety of self-assembled, semimetallic ErSb nanostructures embedded within a GaSb matrix grown by MBE. The size, shape, and orientation of the nanostructures are tunable, controlled solely by the Er concentration during growth. The nanostructures vary from spherical nanoparticles to nanorods and nanowires with tunable orientations. Surface plasmons (SP) arise from the free electrons at the surface of the semimetallic nanostructures and are observed around 110-115 THz. This resonance feature is within the bandgap of the GaSb matrix and can be utilized to engineer the properties of the ErSb:GaSb composite for plasmon-enhanced applications[17-20]. The plasmonic properties are characterized by several techniques including Fourier transform infrared (FTIR) spectroscopy, frequency-domain and time-domain terahertz (THz) measurements, covering an extremely wide range from 100 GHz (3mm) up to 300 THz (1 μm). Strong polarization-sensitive effects are observed for the nanowires embedded in the growth plane. A model based on Maxwell-Garnett (MG) effective medium theory is applied and the effects of size, shape, and orientation of the ErSb nanostructures are well explained using the model over the whole investigated frequency range. Parameters such as carrier concentration in the semimetallic ErSb can be extracted from the model, and deviations from a simple Drude model suggest that quantum size effects (QSE) and inter-grain coupling should be considered. Unlike metal particles that have resonance features in the visible to UV wavelength range, the infrared (IR) SP from the semimetallic nanostructures can extend research on plasmonics[21-23] in the IR and THz regions.

GaSb has been used widely in IR applications, such as lasers and photodetectors. ErSb has attracted interest for making epitaxial contacts to GaSb and superlattice structures with GaSb[3]. The most thermodynamically stable crystal structures for the two materials are different, zincblende and rock-salt for GaSb and ErSb, respectively. Fortunately, ErSb and GaSb share the same face-centered cubic Sb sublattices that allows for a continuous Sb sublattice grown across the ErSb and GaSb interface to make the composite highly coherent. There have been reports on ErSb growth on GaSb[24,25], and an embedded growth mode has been proposed by Schultz[26] et al. However, most of the studies so far are done on submonolayer depositions of ErSb on the surface of GaSb. Generally, incomplete ErSb coverage is needed for nucleation of subsequent defect-free GaSb overgrowth; fully covered ErSb-terminated surfaces cannot easily be overgrown. Here we employ a codeposition process in which ErSb and GaSb are grown simultaneously and continuously. We further develop a method to control the formation and evolution of ErSb nanostructures simply by adapting the Er concentration.

Sample X, a multi-layered structure (Fig. 1a), was grown with increasing Er concentration in order to investigate the growth modes, growth coherency and nanostructure formation of ErSb in GaSb. Unintentionally doped GaSb layers were used to separate the ErSb:GaSb composite layers. Cross-section transmission electron microscopy (TEM) image (Fig. 1b) shows that a variety of nanostructures are formed with different Er concentrations. More nanostructure details are observed along two perpendicular directions [110] (Fig. 1c) and [-110] (Fig. 1d) with higher magnifications. At a concentration of about 1% Er, the solubility limit is reached such that Er and Sb cluster into nanoparticles (Fig. 1, bottom ErSb:GaSb layer). As the Er concentration is increased, the nanoparticles grow in both size and density and a short-range alignment becomes apparent (Fig. 1, second ErSb:GaSb layer from the bottom).

At an Er concentration of ~7%, the ErSb clusters self-align along the growth direction, forming elongated, rod-shaped structures with long-range order (Fig. 1, middle layer). As the Er concentration is further increased to ~10%, the rod-shaped nanoparticles begin to grow continuously, forming nanowires along the growth direction (Fig. 1, second ErSb:GaSb layer from the top). When the Er concentration is increased to ~20%, the nanowire formation rotates by 90 degrees and the wires align horizontally along the <-110> direction (Fig. 1, top ErSb:GaSb layer). No stacking faults are visible through the entire structure indicating excellent crystalline quality of both the ErSb and the GaSb matrix for Er concentration up to 20%. There is no evidence that the growth and nanostructure formation in the top layers are affected by the layers grown underneath, showing that the composites can easily be overgrown.

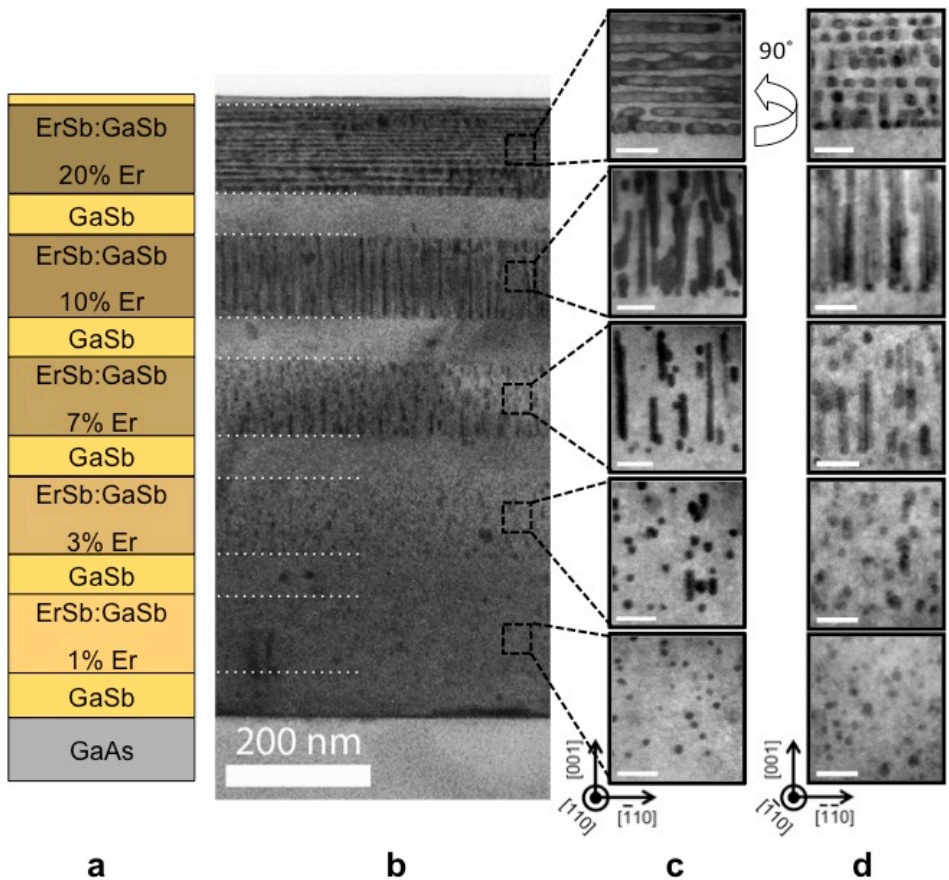

**Figure 1 Cross sectional TEM image of sample X consisting of five layers of ErSb:GaSb with different Er concentrations. a** A schematic illustration of the stack structure of sample X, the nominal Er concentrations are 1%, 3%, 7%, 10%, and 20%, from the bottom to the top. **b** TEM along the [110] zone axis showing an overview of the sample. The layered structure is clearly resolved. **c** Images along [110] of each layer and **d** images along [-110] which is perpendicular to [110] of each layer with higher magnification. **c** and **d** show the formation and evolution

of the ErSb nanostructures from spherical nanoparticles (1% Er), to bigger ErSb nanoparticles with short-range alignment (3% Er), to short ErSb nanorods along the growth direction (7% Er), to almost continuous vertical ErSb nanowires (10% Er), to nanowires embedded in the growth plane (20% Er). All the scale bars in the higher magnification images are 20 nm.

ErSb exhibits a semimetallic band structure[27], similar to what is seen in ErAs[28,29]. This is supported by the observation of a large dc-conductivity ($\sigma_{ErSb}$=35,500 S/cm) in an MBE-grown ErSb film[30]. Therefore, the variety of ErSb nanostructures self-assembled in GaSb provides a material system with large tunability to study the properties of ErSb. In order to investigate the electrical and optical properties of the composites with different nanostructures, three separate ErSb:GaSb samples with 3% (sample A), 10% (sample B) and 20% (sample C) Er concentrations were grown with similar growth conditions but with a nominal thickness of 2 µm. A 2 µm-thick GaSb film was grown as a reference sample (sample D). TEM images of sample B show continuity of the nanowires grown along the growth direction (Fig. 2a) and uniform distribution and ordering of the nanowires (Fig. 2b). A Z-contrast high-angle annular dark-field (HAADF) scanning transmission electron microscopy (STEM) image (Fig. 2c) resolves the octagonal shape of the nanowires with a diameter of ~7 nm. Figure 2d-f show the cross-section HAADF-STEM images of the 20% Er samples. We note in Fig. 2d that a number of faceted nanowires are embedded in the growth plane in a layered manner. In Fig. 2e we observe clear discontinuities in the in-plane nanowires suggesting that these wires are formed from a chain of inter-connected nanorods. The high resolution TEM image Fig. 2f reveals the abrupt interface between the rock-salt ErSb nanowire and the zincblende GaSb matrix indicating excellent overgrowth of the nanorods: no stacking faults are visible at this level.

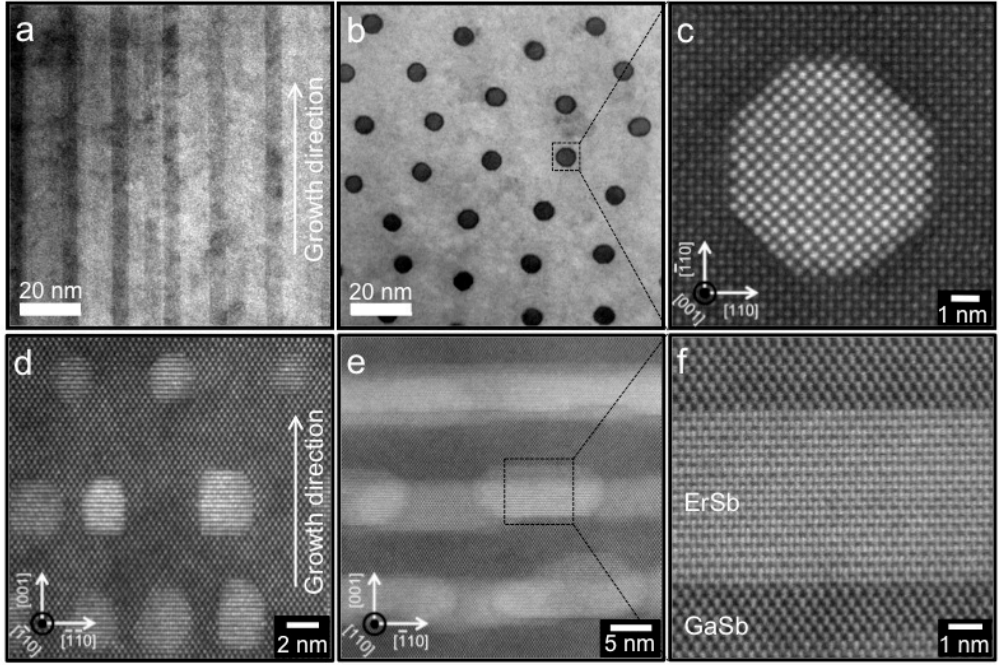

**Figure 2 a-c Images of an ErSb:GaSb sample with 10% Er (sample B), d-f HAADF-STEM images of ErSb:GaSb sample with 20% Er (top layer of sample X and sample C). a** Cross sectional bright field TEM image showing the continuity of highly ordered ErSb nanowires. **b** Plan view bright field TEM image showing the ErSb nanowire array with uniform distribution and small size variation. **c** HAADF-STEM image along the [001] zone axis revealing an octagonal shape of the ErSb nanowire having atomically abrupt interfaces with the GaSb matrix due to the atomic number sensitivity of the image. **d** Image along the [-110] zone axis of the top layer of sample X showing the cross section of the nanowires embedded in-plane with faceting effect. **e** Image along the [110] zone axis of sample C showing the overlapping of the ErSb nanowire grains which forms a possible basis for a capacitive coupling between the nanorods. **f** Image with a high magnification resolving the rock-salt crystal structure of the ErSb nanorod and the zincblende crystal structure of the GaSb matrix that are perfectly coherent with atomically abrupt interfaces. The image also shows the Sb sublattice is continuous across the interface of the GaSb matrix and the ErSb nanorod. Note that it is difficult to resolve the rock-salt structure in **d** due to the overlap of the embedded nanorod with the matrix along the viewing direction.

Electrical dc measurements revealed a much lower conductivity in all samples compared to a continuous ErSb film. We attribute the low dc-conductivity to the discontinuity of the ErSb nanostructures in the current flow direction. The substructure observed in Fig. 2e suggests limited continuous macroscopic current pathways along the in-plane nanowires, confirmed by the low dc-conductivity. In order to further investigate the material properties of the ErSb nanostructures, polarization-resolved spectroscopy is performed at IR and THz frequencies. Using a combination of three different techniques, FTIR, frequency-domain, and time-domain THz spectroscopy, the dielectric tensor of the nanostructured media was measured over a frequency range of 0.1-300 THz. When the light is polarized transverse to the nanowire axis, or when nanoparticles are studied, free carriers in the ErSb will oscillate across the nanostructures in a SP resonance, which couples strongly to incident IR radiation[13,31]. Additionally, IR light will couple to electron transitions in the nanostructures.

Terahertz to IR transmission and reflection measurements were taken at normal incidence with light polarized parallel and perpendicular to the <-110> axis (Fig. 3). For sample A and B, very little polarization dependence was observed, consistent with the isotropy in the growth plane. Additionally, the similarity to the reference sample D at low frequencies (< 20 THz) indicates that capacitive coupling between wires is weak since the wires in sample B are perpendicular to the electric field of the light under normal incidence. By contrast, nanowires embedded in-plane showed a very large polarization contrast (Fig. 3c) over the entire measurement range. This is associated with the large ac-conductivity along the in-plane nanowires which is probed by the light polarized along the <-110> axis. The polarization contrast persists to the THz region measured by frequency- and time-domain THz spectroscopy (Fig. 4a,b). Below 80 THz, sample C behaves strongly like a wire-grid polarizer due to the large ac-conductivity of the in-plane nanowires.

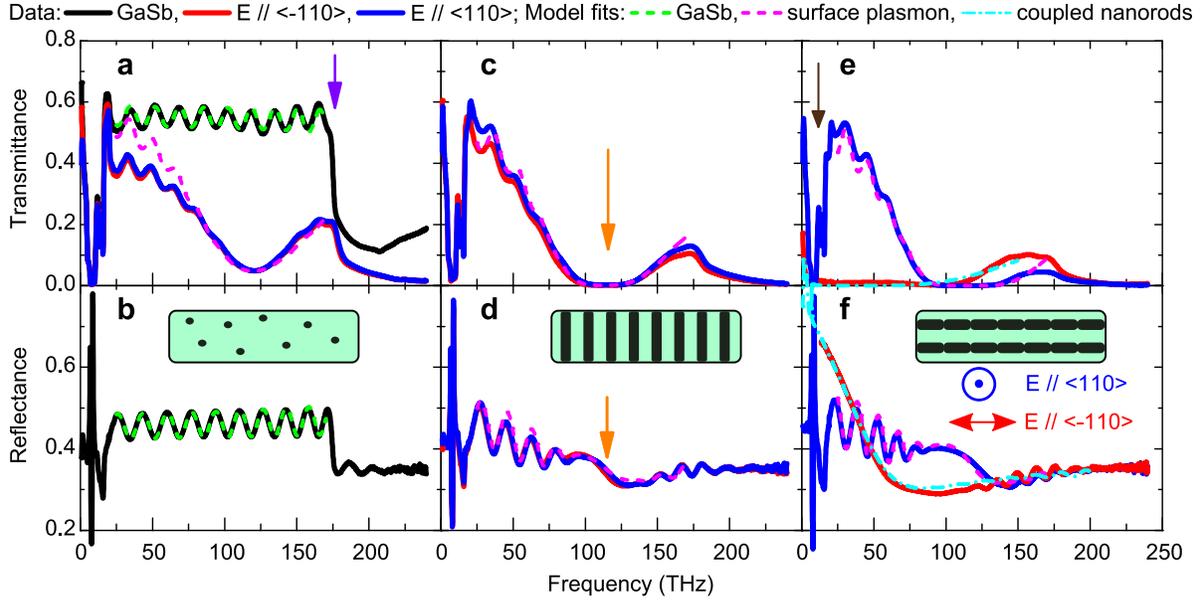

**Figure 3 Transmittance (top row) and reflectance spectra (bottom row) of the three ErSb:GaSb composite samples and the GaSb reference sample.** Solid lines indicate the measured spectra with polarization parallel to the <-110> (red) and <110> (blue) directions. The spectra for the reference 2 μm thick GaSb film are shown in black. Dashed lines indicate model fits. **a** Transmittance and **b** reflectance of sample A where nanoparticles are formed at 3% Er inclusion and of the reference sample D. The oscillations in the GaSb data are due to standing waves within the active layer. A purple arrow indicates the onset of interband absorption in the GaSb. **c** Transmittance and **d** reflectance of sample B, 10% Er, in which the nanowires are oriented along the growth direction. The orange arrows indicate the surface plasmon resonance also seen in the other samples. **e** Transmittance **f** reflectance of sample C, 20% Er, having in-plane nanowires. The electric field orientations of the respective IR/THz polarizations are indicated in **f**, where a large polarization contrast arises due to large ac-conductivity along the nanowires, particularly from the transmission shown in **e**. Phonon absorption in the matrix and substrate are indicated by a brown arrow in **e**.

With light polarized perpendicular to the nanowires in sample C and for both polarizations of light in samples A and B, there is a transmission minimum and a dispersive feature in the reflectance at 110-115 THz. We attribute this resonance to a SP which is discussed in the model below. Oscillations in the IR spectra with a period ~17 THz are quantitatively identified with coherent internal reflections in the films and provide a measure of the film thickness in each sample. Optical phonon absorption in the substrate and matrix can be seen between 3-17 THz. The onset of GaSb interband absorption is visible at 174 THz (~1.72 μm).

The frequency-dependent dielectric response, $\varepsilon(\omega)$, or ac-conductivity, $\sigma(\omega) = -i\omega[\varepsilon(\omega)-\varepsilon_\infty]$ were extracted from model fits to the spectra. Maxwell-Garnett (MG) effective medium theory[32] was employed for the ErSb:GaSb composites and the reflection and transmission coefficients were calculated according to a transfer matrix method. The GaSb optical constants are available in the literature[33] and from measurements of a bare GaSb substrate and the reference sample D. For the ErSb nanostructures we start with the model dielectric function for a Lorentz oscillator:

$$\varepsilon_{ErSb}(\omega) = \varepsilon_\infty + \frac{\omega_p^2}{\omega_0^2 - \omega^2 - i\gamma\omega}, \tag{1}$$

The plasma frequency, $\omega_p^2 = \Sigma_i n_i e^2/m_i \varepsilon_0$, is a sum over electron and hole conduction bands $i$. The scattering of charge carriers in the ErSb is described by the damping factor $\gamma$. The constant $\varepsilon_\infty$ arises from excitations at energies above the measurement range and $\varepsilon_o$ is the vacuum permittivity. A resonant frequency $\omega_0$ is included and interpreted as a QSE or a capacitive coupling, depending on the structure and polarization, and is discussed below. For $\omega_0 = 0$ the Drude model of a metal is recovered. We set $\varepsilon_\infty = 20$ for all samples – this number was obtained with ~25% uncertainty by the fits to sample C where the most ErSb is present. Earlier measurements on elliptical ErAs nanostructures have been treated in a similar fashion[34], with the exclusion of the resonant feature.

First, consider the polarization *along* the in-plane nanowire major axis (<-110>) in sample C. The MG theory reduces to a volumetric average of the nanowires and GaSb matrix dielectric response functions. However, MG theory cannot account for the capacitive coupling across the discontinuities in the wires; therefore we employ an effective circuit model, which takes form of Eq. (1) with the resonant frequency $\omega_0^2 = (LC)^{-1} = \varepsilon_0 \omega_p^2/Ca$. $L$ is the kinetic inductance of the semimetallic ErSb, $C$ is the capacitive coupling along their axis, and $a$ is the average length of the rod-shaped nanowire grains. An accurate measurement of the ErSb plasma frequency is given by the integrated strength of the absorption, independent of *C*.

Second, consider the nanoparticles (sample A) and light polarized *transverse* to the nanowires (samples B and C). Application of MG theory predicts a SP resonance (even for $\omega_0 = 0$) at a frequency related to the ErSb plasma frequency and nanostructure shape. Additionally, to consider QSE, a non-zero $\omega_0$ is included in the nanostructure dielectric function Eq. (1) to approximately account for transitions between quantum confined states separated by energies of order $\hbar\omega_0$, whose presence may be expected given the nanometer dimensions of the nanostructures. Application of MG theory predicts SP absorption at the frequency

$$\omega_{SP}^2 = \omega_0^2 + \frac{(1-\eta)P\omega_p^2}{\varepsilon_m + P(1-\eta)(\varepsilon_\infty - \varepsilon_m)}, \qquad (2)$$

where $\eta$ is the Er volume fraction, $\varepsilon_m$ is the dielectric constant of GaSb, and the depolarization factor *P* is 1/3 for a sphere and 1/2 perpendicular to the axis of a cylinder. From the perspective of MG theory, we treat all nanowires as cylinders. This treatment is supported by the strong inter-grain coupling and by the lack of shift between the two types of nanowire samples (B and C). The nanoparticles are modeled as spheres, although some shape distribution is observed. In this model QSE provides a blue-shift of the SP response of otherwise free carriers.

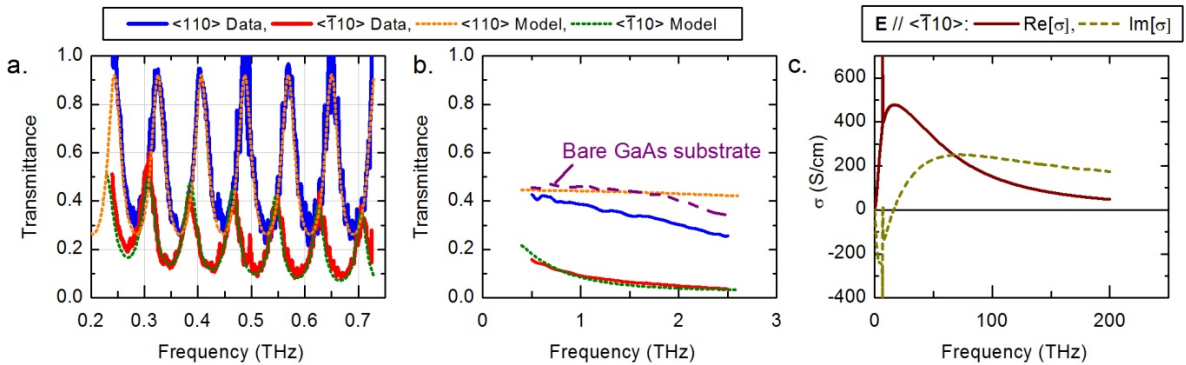

**Figure 4 Frequency-domain (a) and Time-domain (b) THz transmission spectroscopies for sample C having in-plane nanowires.** The polarization is either parallel to the wires (red) or perpendicular (blue). The simultaneous fit to the coupled grain model is shown in green. The model on transmission perpendicular to the wires is shown in

orange. The transmission through a bare GaAs substrate is also shown in **b**. In the frequency domain measurement **a**, an oscillatory spectrum results from standing waves inside the sample. **c** The ac-conductivity of the composite medium in sample C, measured along the <-110> orientation, is from the fits to data in **a** and **b**, and Fig. 3**e** and **f**. The resonance is interpreted as arising from the reactive part of the Drude response of the ErSb and the capacitive coupling of nanowire grains along their alignment direction <-110>.

For light polarized along the in-plane nanowires in sample C, the entire range of data is very well fit by taking Eq. 1 for the ErSb, with the fit parameters shown in Table 1. Still, to achieve the best fit at frequencies below 1 THz, we *additionally* included a comparatively very weak Drude term for the composite medium with a dc-conductivity of 10±5 S/cm and $\gamma_{GaSb}/2\pi = 8±4$ THz, accounting for the doping by Er in GaSb matrix and possible resistive coupling between the nanowire grains. The fits are shown together with the data in Fig. 3e-f and Fig. 4a-b. The ac-conductivity of the composite medium derived from the fit is shown in Fig. 4c. The fit gives an ErSb plasma frequency $\omega_p/2\pi = 540$ THz, close to the value 470 THz calculated from the band structure of the related compound ErAs[29,34]. The transmission increases at frequencies below the *LC* frequency (17 THz) of the nanowires. Nonetheless, the transmission contrast between polarizations remains significant, greater than 50%, down to the 0.15 THz limit of the measurement (Fig. 4a). Taking an average grain length of 15 nm (Fig. 2e), we obtain a reasonable value for the inter-grain capacitance equal to $5.1 \times 10^{-18}$ F, corresponding to an inter-grain distance in the range of ~1 nm according to a simple plate capacitor estimate. Thus this model can reconcile the low dc-conductivity with a semimetallic band structure. In the case of a continuous structure with $\omega_0 = 0$, the scattering rate and plasma frequency obtained from the fit would give a dc-conductivity of 2,400 S/cm. This suggests the surface of the nanostructures is a dominant source of carrier scattering. For the nanoparticles, and the nanowires with polarization perpendicular to their alignment direction, the absorption spectra are well fit by the model for SP absorption, as indicated in Fig. 3. The plasma frequencies in Table 1 are in reasonable agreement to those obtained from the measurement with polarization along in-plane nanowires, though systematically higher. Additionally, intersubband transition energies $\hbar\omega_0 \sim$ 150-350 meV are plausible given the dimensions of the nanostructures[28]. The scattering rates are only weakly dependent of the particle type or IR polarization. For sample C at frequencies of 1-2.5 THz a small amount of absorption is observed for polarization perpendicular to the nanowires which is not accounted for in the model (Fig. 4b).

Mixing of intersubband transitions and SP has been well documented in the absorption spectra of quantum wells[35] and quantum dots[36]. In the present study, attempts to fit the SP absorption while constraining $\omega_0 = 0$ to neglect QSE consistently overestimated the strength of the absorption as seen in both the reflectance and transmittance and produced comparatively large plasma frequency, corresponding to up to 2 fold increase in the ErSb carrier density. An important *qualitative* feature also emerges in comparing the nanoparticles to the nanowires. In the absence of QSE, one would expect the nanoparticle SP frequency to be ~ 18% lower than the nanowires due to the lower depolarization factor $P$ for a sphere, in contrast to the results. This suggests a careful analysis[37] of quantum effects in these nanostructures is useful. Finally, we note that the neglected coupling between adjacent nanowires (in the <110> orientation) could lead to a reduction of the SP frequency[37] at higher Er inclusion.

In summary, a self-assembled semimetal/semiconductor composite material system ErSb:GaSb is grown by MBE. The size, shape and orientations of the ErSb nanosructures can be controlled solely by the Er concentration. The semimetallic behavior of the ErSb is corroborated by the observation of SP oscillations. A simple model accounts for the polarization dependent plasmonic properties of the nanostructures over three frequency decades. Additionally, a large ac-conductivity along in-plane oriented nanowires leads to a large polarization contrast like wire-grid polarizers up to ~ 80 THz.

These structures may present an opportunity to improve the polarization characteristics of IR and THz quantum cascade lasers since they can be grown below and above the active layers without deteriorating the crystal quality. The ability to control the size, shape and dimensionality of ErSb nanostructures can lead to the development of grown passive and active optical devices in the IR and THz regions of the spectrum. Additionally, these structures provide anisotropic electron and phonon transport that can be exploited to improve the efficiency of thermoelectric power conversion. The self-assembly of this material system provides a route to grow and engineer complex arrays of metallic nanostructures in which excitation and manipulation of SP can be realized. This may provide a potential opportunity to the ErSb:GaSb or ErSb:Ga(In)Sb material systems to be applied in the fast grown field of plasmonics. Notably, the TEM images indicate that the vertical nanowires are considerably more continuous than the horizontal nanowires. Future efforts aimed at increasing the continuity of horizontal nanowires may lead to decreased transmission and increased polarization contrast at frequencies near 1 THz. The perfect octagonal shape of the ErSb nanowires embedded in GaSb provide a material system to study equilibrium crystal formation by Wulff construction[38]. Furthermore, a highly ordered composite system with embedded semimetallic nanowires can be applied to other rare-earth-V based material systems. With more tunability in the optical properties such as permittivity, anisotropy and magnetic properties, we expect a potential application of the rare-earth-V compounds as self-assembled nanoplasmonic metamaterials[39,40].

| Sample | Film thickness ($\mu m$) (from fit) | IR Polarization | $\omega_p/2\pi$ (THz) | $\omega_0/2\pi$ (THz) | $\gamma/2\pi$ (THz) | $\omega_{SP}/2\pi$ (THz) |
|---|---|---|---|---|---|---|
| A (3%) | 2.3 ± 0.1 | Either | 720 ± 100 | 62 ± 30 | 65 ± 20 | 130 ± 5 |
| B (10%) | 2.0 ± 0.1 | <110> | 610 ± 50 | 51 ± 20 | 45 ± 10 | 117 ± 3 |
| | | <-110> | 640 ± 50 | 36 ± 20 | 45 ± 10 | 118 ± 3 |
| C (20%) | 2.3 ± 0.1 | <110> | 500 ± 60 | 82 ± 20 | 40 ± 5 | 112 ± 3 |
| | | <-110> | 543 ± 14 | 17 ± 1 | 66 ± 6 | 117 ± 3 |

**Table 1 Fit parameters for the ErSb nanowires and nanoparticles (Eq. 2).** The resonant frequencies are interpreted as being characteristic of quantum transitions, except for sample C polarized along <-110> where $\omega_0 = (LC)^{-1/2}$ in a circuit model for the in-plane nanowires. The SP frequencies are given according to Eq. (2); the peak absorption in Fig. 3 is at a lower frequency due to damping.

**Methods summary**

Growth of ErSb:GaSb composite

All the ErSb:GaSb samples were grown by solid-source MBE on double-side polished semi-insulating (100) oriented GaAs substrates. A 200 nm thick GaAs buffer layer for smoothing the surface is followed by a 60 nm thick GaSb buffer layer for stress relaxation between GaAs and GaSb resulting from a lattice mismatch of 7.8%. The following ErSb:GaSb composites layers were grown during a codeposition process.

Characterization

The bright-field TEM imaging was performed by TEM Analysis Service Lab using a Philips 420 TEM operated at 120 kV. STEM imaging were carried out using a FEI Titan STEM/TEM equipped with a field-emission electron gun operated at 300 kV, at a Materials Research Laboratory facility at UCSB. The atomic number sensitive HAADF-STEM images were recorded using an annular dark-field detector. Plan-view and cross-section TEM samples were prepared by a FEI Helios focus ion beam (FIB) system.

DC conductivity measurements were performed on a standard Hall bar geometry fabricated along both <110> and <-110> orientations on the samples by photolithography.

Polarization-resolved measurements and modeling

Transmittance and reflectance measurements were made using three different spectrometers. The range of frequencies 0.15-0.73 THz was measured using continuous wave near normal transmission spectroscopy in which a Rhode-Schwartz vector network analyzer was connected to Virginia Diodes frequency extenders. An oscillatory transmission spectrum results due to etalon effects in the sample; the amplitude and phase of the oscillations are affected by the ac-conductivity of the nanostructured films. To cover the range 0.5-2.5 THz, we employed a THz time domain spectrometer using an asynchronous optical sampling technique. Since internal reflections are delayed in the time domain, this technique measures the single pass transmission coefficient. Finally, the range 2-270 THz was measured in a Bruker 66v/S Fourier transform IR spectrometer, operated in vacuum, in both near-normal (~10°) reflectance and normal transmittance geometries. The throughput was carefully adjusted to minimize the effect of detector non-linearity. All spectroscopic measurements were made with polarization along either the <110> or the <-110> orientations of the substrate and at a temperature 22° ± 3° C.

**Supplemental information**

**Growth**

The ErSb:GaSb samples were grown on 0.5 mm thick, semi-insulating (100) GaAs substrates by solid-source MBE using a Varian Gen II system equipped with arsenic and antimony valved crackers. A GaAs substrate is selected for several reasons: its semi-insulating property allows us to perform electrical measurement directly on the film-of-interest without removing the substrate; its wider bandgap allows us to observe the absorption that occurs within the bandgap; and its low intrinsic carrier density reduces free carrier absorption in the spectrum. The disadvantage of using GaAs substrate is to deal with the 7.8% lattice mismatch between the film-of-interest and the substrate. After the native oxide was thermally desorbed from the surface of the GaAs substrate, a 200 nm thick GaAs buffer layer for smoothing the surface was followed by a 60 nm thick GaSb buffer layer for stress relaxation between GaAs and GaSb. For Sample A and B, the GaSb buffer layer was eliminated because the same stress can be relaxed by the Er-containing GaSb layer and the ErSb nanostructure formation is not affected, which was confirmed by separate TEM study. $As_2$ and $Sb_2$ were used for the arsenide and antimonide growths. The Er source is a solid-source effusion cell operated at temperatures between 900°C to 1200°C for varying the Er concentrations. The growth temperatures of the GaAs buffer layer and the GaSb layers (with or without ErSb) were kept at 600°C and 500-530°C, respectively. In-situ reflection high energy electron diffraction (RHEED) and ex-situ atomic force microscopy (AFM) studies showed that the growth mode of the ErSb:GaSb was step-flow in the applied growth temperature range. To study the formation mechanism of the ErSb nanostructure, 60 nm thick unintentional doped GaSb layers were sandwiched between the ErSb:GaSb layers in Sample X for interface identification. All the ErSb:GaSb layers were overgrown by the GaSb intermediate layers without indication of crystalline defects. All the samples and related ErSb nanostructures formation have been reproduced on several MBE systems. X-ray diffraction (XRD) spectra (as shown in Figure 1) on similar samples with Er concentration up to 10% do not show any phase separation on the ErSb:GaSb films indicating a perfect coherence of the composites. There is also no broadening in the Er-containing films comparing with the reference GaSb film.

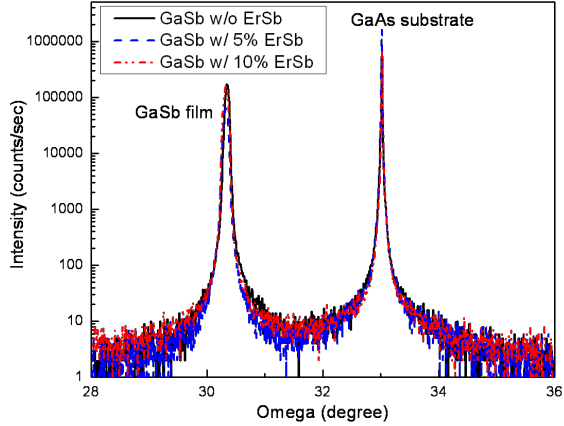

**Figure 1 XRD spectra of three samples: GaSb reference, 5% ErSb and 10% ErSb grown on GaAs substrate.** The peak separation between the GaAs substrate and the GaSb samples corresponds to a lattice mismatch of ~7.8%, showing a fully relaxation of the epi-films on the substrate.

### Direct-current conductivity

Room temperature resistivity measurements of the nanostructured materials were made using a four-point probe station on a standard Hall-bar geometry. The Hall-bar devices were prepared using standard lithography techniques. Metallization patterns were defined in AZ5214 by optical contact lithography and a metal contact stack of 50 nm Ti under 500 nm Au was deposited by E-beam. After metallization and lift-off, mesa patterns along both <110> and <-110> orientations were defined in AZ4210 and the GaSb was wet-etched using a 100:1:100 volume ratio of $HCl:H_2O_2:H_2O$ which selectively etches GaSb over GaAs. No annealing step was used. The dc-conductivities of the 10% and 20% samples along <110> and <-110> orientations are listed in Table I. The dc-conductivity from a GaSb reference sample is also listed.

| Sample | $\sigma_{<110>}$ (S/cm) | $\sigma_{<-110>}$ (S/cm) |
|---|---|---|
| 10% Er (sample B) | 2.1 ± 0.1 | 2.3 ± 0.1 |
| 20% Er (sample C) | 0.85 ± 0.1 | 3.1 ± 0.2 |
| GaSb reference (sample D) | 4.0 | 4.0 |

**Table I. Conductivity measured at dc.** An anisotropy is observed in the 20% sample with in-plane nanorods, however upon comparison to the 10% sample with vertical nanowires it appears that anisotropic scattering in the GaSb, rather than dc conduction in the wires, is responsible for the effect.

**Continuous Wave THz Transmission**

The range of frequencies 0.15-0.73 THz was measured using continuous wave near normal transmission spectroscopy in which a Rohde-Schwartz vector network analyzer (VNA) was connected to Virginia Diodes frequency extenders. Corrugated horn antennas were used for coupling the highly polarized THz radiation from and to free space. The sample was mounted on a rotator between closely spaced sender and receiver horns. The setup is illustrated in Figure 2. The high coherence of the continuous wave system resulted in a series of unwanted standing waves, such as those between the sender and receiver, or the horns and facets of the sample. Due to the relatively large length between these elements, the oscillations are below 2 GHz; therefore, we smoothed the experimental data sets by averaging over 2.5 GHz, retaining only the 82 GHz oscillations which originate from the standing wave inside the substrate. The ac-conductivity of the nanorods in the 20% ErSb sample was obtained from a strong shift in amplitude and phase of these oscillations upon rotation of the sample. The standing wave pattern therefore strongly increases the accuracy of the fitted dielectric function. Further information on the system is available on the website of Institute for Terahertz Science and Technology at UCSB (http://www.itst.ucsb.edu/vnavdi.html).

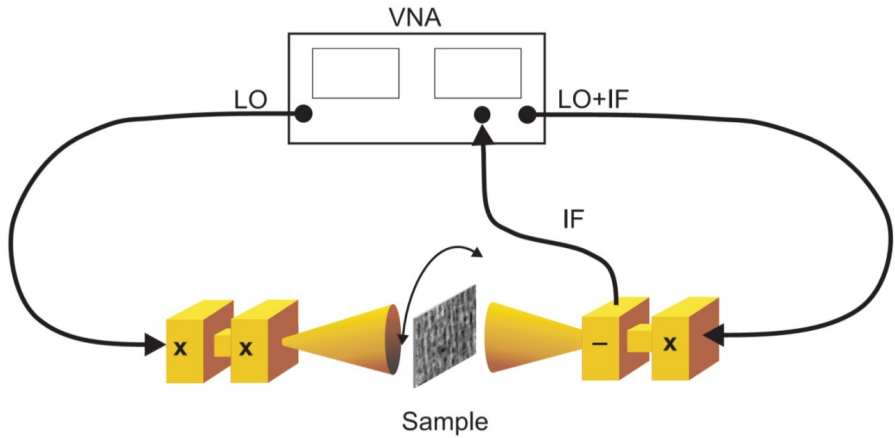

**Figure 2 Illustration of the continuous wave THz spectrometer.**

## THz-Time Domain

A key advantage of this measurement is that internal reflections in the substrate are sufficiently time-delayed to be excluded from the analysis, providing a direct, phase sensitive measurement of the single-pass transmission coefficient. An asynchronous optical sampling (ASOPS) method was used to perform the spectroscopy. In this setup, two 1560 nm IR lasers (Menlo Systems) with 250 MHz repetition rate were detuned from each other in pulse rate by 2.5 KHz to produce a variable time delay between pump and probe pulses. Both the pump and the probe were frequency doubled to produce 780 nm pulses at 150 mW and 500 mW respectively. The THz radiation was produced by pumping a LT-GaAs photoconductive antenna (PCA) and afterwards transmission through the sample was measured by mixing the transmitted field with the NIR probe in a birefringent ZnTe detection crystal. The polarized probe pulse is then split into two components by a polarization beam-splitter and measured with balanced photodiodes. The differential amplitude in the two polarization channels is proportional to the THz electric field. This process is depicted in Figure 3. The power spectrum of the THz pulse can then be obtained by Fourier transforming the measured time evolution of the THz electric field. The THz emitted from the antenna is highly polarized, and the instrument has bandwidth from 0.25 – 2.75 THz with peak power at 1 THz.

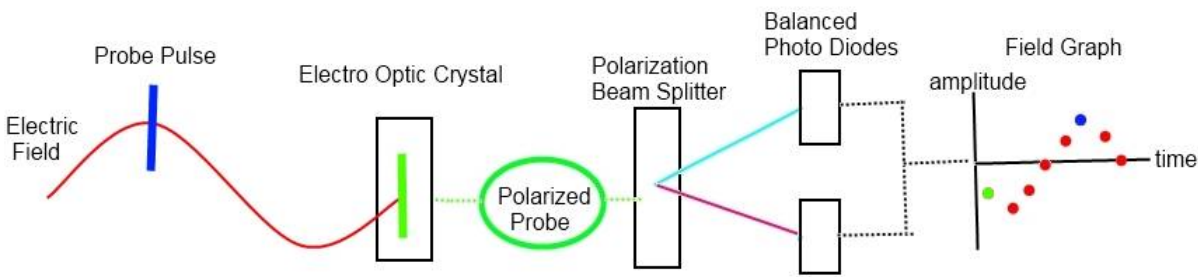

**Figure 3** The relative electric field amplitude at the position of the probe is detected by sending the probe and the field into an electro-optic Crystal (ZnTe) and splitting the resulting elliptically polarized pulse onto a pair of balanced photo diodes. The measured signal difference between the polarizations is directly related to the amplitude of the THz electric field.

Polarization sensitive measurements were made by rotating the sample about the axis of THz propagation on a mount with a 5 mm opening for transmission of the 1-2 mm diameter focus of the THz pulse. All measurements were performed in a purged $N_2$ environment to minimize water absorption.

**FTIR reflectance and transmittance**

Measurements were made in a Bruker 66v/S Fourier Transform IR spectrometer (FTIR), operated in vacuum. Far-IR spectra were taken using a Hg lamp, Ge coated Mylar beamsplitter, and liquid He cooled bolometer. Mid-IR measurements employed a globar source, KBr beamsplitter, and either a HgCdTe (MCT) or pyrolectric DTGS detector. Extreme care was taken to minimize the influence of the MCT detector nonlinearity by limiting throughput, optical filtering, and comparison to the highly linear but low sensitivity DTGS. Samples were mounted on a 5 mm travel translation stage connected to a vacuum feed-through allowing repeatable interchange of samples without breaking vacuum. Reflectivity measurements were made at ~10° incidence and referenced to a 3μm Au film on a GaAs substrate. Transmission measurements were referenced to the open path spectrum using a gap between samples; we accounted for slight lensing and internal reflections in the GaAs substrate. Polarization sensitivity was achieved by placing one of several rotatable polarizers in front of the sample – a free standing far IR wire-grid, a mid IR wire grid on a KRS-5 substrate, and a near-IR thin film polarizer.

**Multilayer model and Maxwell Garnett Theory**

The optical constants of the nanostructures were extracted by fitting the measurements to a multi-layer model based on a transfer matrix solution to Maxwell's equations. For the VNA measurement the coherent internal reflections in the GaAs substrate were also included in the transfer matrix solution. The nanostructured films were treated as an effective medium with a diagonal dielectric tensor $\varepsilon_{eff,j}$, where $j$ denotes a principle axis of the nanostructures. By assuming that capacitive coupling between adjacent nanowires/spheres is negligible, and noting that dimensions are much smaller than the estimated ErSb skin depth, we can apply the quasi-static Maxwell-Garnett effective medium theory for ellipsoidal inclusions aligned axially in a matrix. A nanowire can be considered in the limit that the major axis of the ellipsoid is infinite. (We expect minimal error from approximating the wire cross-section as spherical.) Given dielectric functions $\varepsilon_{inc}$ and $\varepsilon_m$ for the inclusions and the GaSb matrix, respectively, the dielectric tensor of the MG effective medium is given by

$$\varepsilon_{eff,j} = \varepsilon_m + \frac{\eta \varepsilon_m (\varepsilon_{inc} - \varepsilon_m)}{\varepsilon_m + P_j(1-\eta)(\varepsilon_{inc} - \varepsilon_m)}. \tag{1}$$

The volume density $\eta$ of the ErSb inclusions is known from the growth. The shape of the nanostructures enters through the depolarization factors $P_j$: equal to 1/3 for a sphere, 1/2 along the radius of a cylinder, and zero along the axis of an infinite cylinder. In the latter case, the effective medium from eq. (1) reduces to a simple volumetric average

$$\varepsilon_{eff,parallel} = \eta \varepsilon_{wire,parallel} + (1-\eta)\varepsilon_{matrix}. \tag{2}$$

We apply eq. (2) to the analysis of the horizontal nanowires in the 20% ErSb samples.

**Surface plasmon and quantum size effect**

When the nanoparticle dimensions are comparable to a Fermi wavelength, the nanoparticle dielectric function may no longer be described by a free carrier Drude model, but rather by a series of transitions between discrete states,

$$\varepsilon_{Q.C.} = \omega_p^2 \times \sum_j \frac{f_j^2}{\omega_j^2 - \omega^2 - i\gamma_j\omega}, \tag{3}$$

where the sum runs over all pairs of states separated by energy $\hbar\omega_j$. Hard-wall boundary condition calculations for similarly dimensioned ErAs nanoparticles[1] and quantum wells[2] predict confinement energies similar to or larger the Fermi energy as measured from the bottom (top) of the electron (hole) band, predicting a semimetal to semiconductor transition which was not observed. Subsequent calculations which include Thomas-Fermi screening and more realistic boundary conditions[1] give a lower ground state energy, explaining the presence of charge carriers; nonetheless it is reasonable to expect transitions at energies near 50-100 THz, as depicted in Figure 4.

For simplicity we consider the case of a single dominant intersubband transition with frequency $\omega_0$. As discussed in the main text, we can model the dielectric response of a quantum well with a single intersubband transition using the Lorentzian profile

$$\varepsilon_{ErSb}(\omega) = \varepsilon_\infty + \frac{\omega_p^2}{\omega_0^2 - \omega^2 - i\gamma\omega}. \tag{4}$$

The resulting MG *effective medium* also has a Lorentzian profile, up to the weak dispersion in the GaSb matrix:

$$\varepsilon_{eff}(\omega) = \varepsilon_{\infty,eff} + \frac{\omega_{p,eff}^2}{\omega_{SP}^2 - \omega^2 - i\gamma\omega}. \tag{5}$$

The resonant frequency of the effective medium can be thought of as a surface plasmon blue shifted by the intersubband spacing:

$$\omega_{SP}^2 = \omega_0^2 + \frac{(1-\eta)P\omega_p^2}{\varepsilon_m + P(1-\eta)(\varepsilon_\infty - \varepsilon_m)}. \tag{6}$$

The strength of the surface plasmon absorption is given by

$$\omega_{p,eff}^2 = \eta\omega_p^2 + \left(\frac{\varepsilon_m}{\varepsilon_m + P(1-\eta)(\varepsilon_\infty - \varepsilon_m)}\right)^2, \tag{7}$$

independent of $\omega_0$. The high frequency dielectric constant $\varepsilon_{\infty,eff}$ is nearly equal to the matrix dielectric constant $\varepsilon_m$.

**Dielectric tensors**

From the model fits, we obtain the dielectric functions of the effective medium as well as the ErSb. These are summarized in Fig. 4, together with cartoon illustrations of the model. For the appropriate polarization (Fig 4a), the effective medium a surface plasmon is observed, which may couple to quantum transitions due to confinement if they are present. The dielectric function in eq.(5) of the effective media are shown in Fig. 4(b), with a resonant frequency given by eq. (6). The dielectric response function in eq. (4) of the ErSb nanostructures is shown in Fig 4(c), where the resonance has been interpreted as an approximate description of quantum size effects.

For polarization along in-plane nanowires, the effective circuit model in Fig. 4(d) was applied. As discussed in the main text, this model takes the form of equation (4) above, expect with a resonant frequency $\omega_0 = (LC)^{-1/2}$. The LC resonance is revealed in a peak in the dielectric response of the composite medium, shown in Fig. 4(e-f). The capacitance arises from coupling between adjacent rod-shaped grains in the nanowire, and the inductance is the kinetic inductance of the ErSb. A sharp feature at 7 THz is an optical phonon in GaSb. At frequencies below 1 THz, a sudden rise in the imaginary part of the dielectric function results from the small Drude term used to account for the contributions of the GaSb matrix. Measurements of additional 10% and 20% samples were performed with highly similar results. We note that the 17 THz resonant frequency observed for light polarized along the nanowires is much too small to be attributed to a surface plasmon resonance of *decoupled* nanostructures as was done for the resonances observed in the transverse polarization.

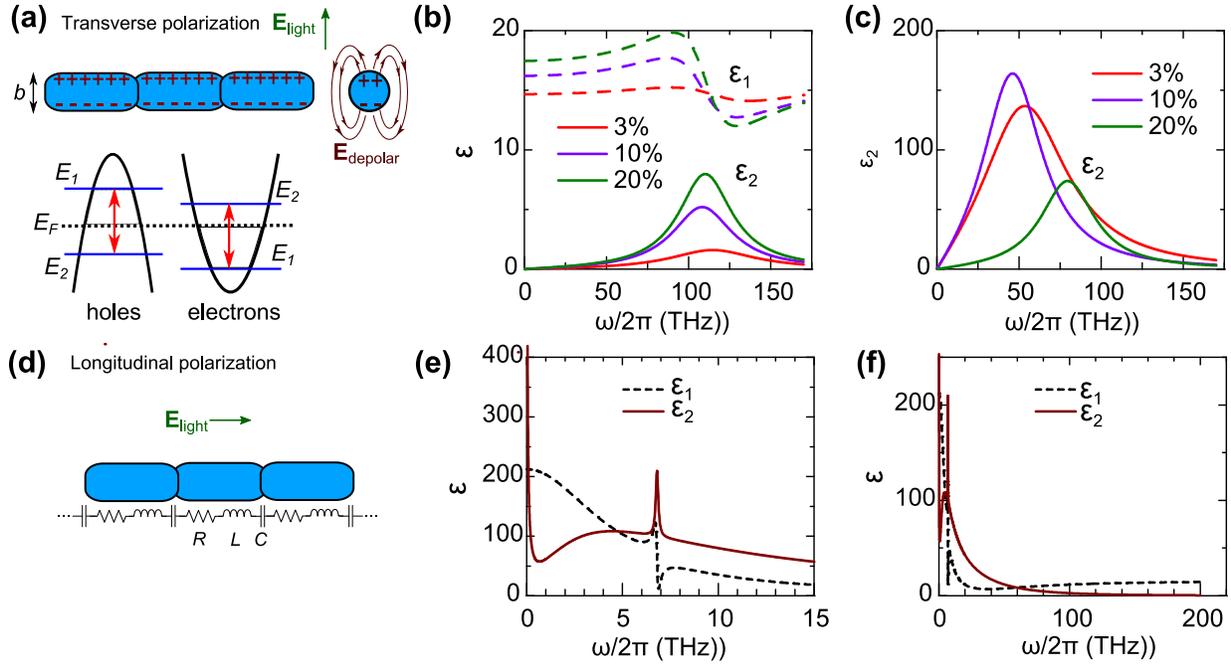

**Figure 4 (a)** The resonant response of the particles and the nanowires with polarization perpendicular to their axis arises from depolarizing electric fields and transitions between quantum confined states. **(b)** Effective medium in samples with various ErSb concentration. **(c)** ErSb dielectric functions used as inputs to the MG theory. The resonances are interpreted loosely as characteristic of intersubband spacings. **(d)** Circuit model for the resonant response when the polarization is parallel to the wire axis **(e)** Effective medium for light polarized along the nanowires in the 20% Er sample. The 17 THz $LC$ resonance is exhibited in the broad peak in Im[$\varepsilon$]; the maximum in Im[$\varepsilon$] occurs at 5 THz due to the strong damping. The GaSb phonon and Drude response appear as sharp features at 7 THz and dc respectively. **(f)** Same as (e) over a wider energy range.